# On linear electromagnetic constitutive laws that define almost-complex structures

**D.H. Delphenich**[*]

Physics Department, Bethany College, Lindsborg, KS 67456, USA

It is shown that not all linear electromagnetic constitutive laws will define almost-complex structures on the bundle of 2-forms on the spacetime manifold when composed with the Poincaré duality isomorphism, but only a restricted class of them that includes linear spatially isotropic and some bi-isotropic constitutive laws. Although this result does not trivialize the formulation of the basic equations of pre-metric electromagnetism, it does affect their reduction to metric electromagnetism by its effect on the types of media that are reducible, and possibly its effect on the way that such media support the propagation of electromagnetic waves.

## 1 Introduction

In conventional Maxwellian electromagnetism, one assumes that the spacetime manifold $M$ has a Lorentzian structure, which is defined by a metric tensor field $g$ that is of globally hyperbolic normal type. It was first observed by Kottler [1] and Cartan [2], and developed further by Van Dantzig [3], that the only manner by which the metric entered into the Maxwell equations:

$$dF = 0, \qquad \delta F = J, \qquad (1.1)$$

was by way of the Hodge duality isomorphism *, which requires the introduction of a metric for its definition, and which is required for the definition of the codifferential operator $\delta$, whose action on $k$-forms on an $n$-dimensional manifold is defined by:

$$\delta = (-1)^{n(n-k)+1} *d* \qquad (1.2)$$

in the Lorentzian case.

A basic property of the Hodge * isomorphism for a four-dimensional Lorentzian manifold $M$ is that it defines an *almost-complex structure* on the real vector bundle $\Lambda^2(M)$ of 2-forms on $M$. That is, the map $*: \Lambda^2(M) \to \Lambda^2(M)$ has the properties that it takes each fiber $\Lambda^2_x(M)$ to itself in a linear and invertible way and:

$$*^2 = -I. \qquad (1.3)$$

---

[*] E-mail: delphenichd@bethanylb.edu

Hence, the action of * on each fiber is completely analogous to the multiplication of complex vectors by $i = \sqrt{-1}$.

To clarify the terminology, one says that * defines a *complex structure* on each fiber, which is a vector space, and an almost-complex structure on the bundle $\Lambda^2(M)$, which defines a module of sections over the ring of smooth functions on $M$. The use of the prefix "almost" refers to the fact that when the vector bundle in question is the tangent bundle to an even-dimensional manifold, it is not always true that the complex structures on the tangent spaces can be integrated into an atlas of complex coordinate systems. However, since the dimension of the fibers of $\Lambda^2(M)$ does not equal the dimension of $M$, in general, integrability is not an issue in the present analysis.

In pre-metric electromagnetism, one replaces the Hodge * with the composition $\tilde{\kappa} = \kappa \cdot \#$ of the Poincaré duality isomorphism #: $\Lambda_2(M) \to \Lambda^2(M)$ that one obtains from a choice of unit-volume element $V$ and a linear constitutive isomorphism $\kappa$: $\Lambda_2(M) \to \Lambda^2(M)$.

Naturally, in order for the isomorphism $\tilde{\kappa}$ to substitute for the * isomorphism, one must have that its square is proportional to $-I$. As we shall show, this implies necessary and sufficient conditions on the component submatrices of $\tilde{\kappa}$ when one chooses a basis for $\Lambda^2(M)$. Hence, one must observe that pre-metric electromagnetism is more general than the metric form. Indeed, since the introduction of * is actually equivalent to the introduction of a conformal class of Lorentzian metrics on $T(M)$, it is clear that the reduction in scope is related to the possibility that the electromagnetic medium in question supports the propagation of electromagnetic waves.

One sees that if one does not wish to start with a Lorentzian structure on spacetime then if the choice of linear electromagnetic constitutive law is to effectively substitute for the Hodge * one must settle for a restricted class of constitutive laws. In particular, although spatially isotropic and bi-isotropic laws permit such a reduction, such a physically reasonable law as the one for an anisotropic dielectric does not. Since an almost-complex structure on the bundle of 2-forms is equivalent to a choice of a conformal class of Lorentzian metrics, one sees that the existence of an almost-complex structure is essential for the definition of the light cones that are associated with the propagation of electromagnetic waves.

In the first section, we briefly summarize the notion of Poincaré duality and introduce some notations. In the second section, we summarize the essential concepts from pre-metric electromagnetism. Then, we discuss the phenomenology of electromagnetic constitutive laws in general. Next, we compose the two isomorphisms $\kappa$ and #, and derive the necessary and sufficient conditions for the square of the resulting isomorphism to be proportional to $-I$. Finally, we discuss the physical nature of these conditions on $\tilde{\kappa}$ that we derived.

## 2 Poincaré duality

Suppose that the manifold $M$ – or rather, its tangent bundle – is orientable and that we have chosen a unit-volume element $V \in \Lambda^4(M)$ on $M$. For instance, when $M = \mathbb{R}^4$ is given a choice of frame $\{\mathbf{e}_\mu, \mu = 0, 1, 2, 3\}$ we can define a unit-volume element by:

$$V = \theta^0 \wedge \theta^1 \wedge \theta^2 \wedge \theta^3 = \frac{1}{4!} \varepsilon_{\kappa\lambda\mu\nu} \theta^\kappa \wedge \theta^\lambda \wedge \theta^\mu \wedge \theta^\nu. \tag{2.1}$$

in which $\{\theta^\mu, \mu = 0, 1, 2, 3\}$ is the reciprocal coframe for $\mathbb{R}^{*4}$. Hence, by definition:

$$\theta^\mu(\mathbf{e}_\nu) = \delta^\mu_\nu. \tag{2.2}$$

Since the vector bundles $\Lambda^2(\mathbb{R}^4)$ and $\Lambda_2(\mathbb{R}^4)$ are trivial, we shall deal with the vector spaces $A^2(\mathbb{R}^4) = \mathbb{R}^{*4} \wedge \mathbb{R}^{*4}$ and $A_2(\mathbb{R}^4) = \mathbb{R}^4 \wedge \mathbb{R}^4$, which are the fibers of those bundles, respectively. The frame $\mathbf{e}_\mu$ and its coframe then define frames for $A_2(\mathbb{R}^4)$ and $A^2(\mathbb{R}^4)$ by way of:

$$\begin{aligned} \mathbf{b}_i &= \mathbf{e}_0 \wedge \mathbf{e}_i, & i &= 1, 2, 3, & \mathbf{b}_4 &= \mathbf{e}_2 \wedge \mathbf{e}_3, \; \mathbf{b}_5 = \mathbf{e}_3 \wedge \mathbf{e}_1, \; \mathbf{b}_6 = \mathbf{e}_1 \wedge \mathbf{e}_2, \end{aligned} \tag{2.3a}$$
$$b^i = \theta^0 \wedge \theta^i, \quad i = 1, 2, 3, \quad b^4 = \theta^2 \wedge \theta^3, \; b^5 = \theta^3 \wedge \theta^1, \; b^6 = \theta^1 \wedge \theta^2. \tag{2.3b}$$

Hence, the coframe $\{b^I, I = 1, \ldots, 6\}$ is also reciprocal to the frame $\{\mathbf{b}_I, I = 1, \ldots, 6\}$ [1].

The Poincaré duality isomorphism then takes the form:

$$\#: \Lambda_2(M) \to \Lambda^2(M), \mathbf{a} \mapsto i_\mathbf{a} V. \tag{2.4}$$

In particular, a decomposable bivector field $\mathbf{u} \wedge \mathbf{v} = \frac{1}{2}(u^\mu v^\nu - u^\nu v^\mu)\, \mathbf{e}_\mu \wedge \mathbf{e}_\nu$ goes to:

$$\#(\mathbf{u} \wedge \mathbf{v}) = i_\mathbf{v} i_\mathbf{u} V = \frac{1}{2} \varepsilon_{\kappa\lambda\mu\nu} u^\kappa v^\lambda \theta^\mu \wedge \theta^\nu. \tag{2.5}$$

The Poincaré duals of the frame elements $\mathbf{b}_I$ can be expressed concisely as:

$$\#\mathbf{b}_i = b^{i+3}, \qquad \#\mathbf{b}_{i+3} = b^i, \tag{2.6}$$

or, in matrix form:

$$[\#]_{IJ} = \begin{bmatrix} 0 & I \\ \hline I & 0 \end{bmatrix}. \tag{2.7}$$

One sees that the duality at issue is the projective-geometric notion that one can define the same 2-plane in $\mathbb{R}^4$ by way of either a pair of linearly independent vectors in the plane that span that plane or a pair of linearly independent covectors that each annihilate it. The former pair gives a non-zero bivector that represents the plane and the latter gives a non-zero 2-form. Moreover, the bivector and 2-form are both defined only up to a non-zero scalar multiple.

---

[1] Unless otherwise stated, from now on, we adopt the convention that Greek characters range from 0 to 3, upper-case Latin characters range from 1 to 6, and lower-case Latin characters range from 1 to 3.

It is important to observe that when one makes a linear invertible change of frame from $\mathbf{e}_\mu$ to $A_\mu^\nu \mathbf{e}_\nu$, although the components $b^{\mu\nu}$ of a bivector $\mathbf{a} = \tfrac{1}{2} a^{\mu\nu} \mathbf{e}_\mu \wedge \mathbf{e}_\nu$ go to $\hat{A}_\kappa^\mu \hat{A}_\lambda^\nu a^{\kappa\lambda}$, in which the tilde denotes the matrix inverse, because one must also transform the components of $V$ by $\det A$, the components $\#a^{\mu\nu}$ of $\#\mathbf{a}$ transform to $(\det A)\hat{A}_\kappa^\mu \hat{A}_\lambda^\nu \# a^{\kappa\lambda}$. Hence, although this gives a tensorial representation of $SL(4; \mathbb{R})$, for which $\det A = 1$ in any case, it does not give a tensorial representation of $GL(4; \mathbb{R})$. This property of dual 2-forms is variously referred to as saying that they are "twisted" 2-forms, "tensor densities of weight 1," or "pseudo-tensors." This situation is analogous to the way that the cross-product of two 3-vectors is referred to as an "axial" vector, whereas the elements of $\mathbb{R}^3$ are "polar" vectors. Indeed, for three dimensions, Poincaré duality takes vectors to 2-forms, which then represent the axial vectors.

## 3 Pre-metric electromagnetism

It is best to regard the Hodge duality isomorphism $*$ as the composition of two isomorphisms, the isomorphism $g \wedge g \colon \Lambda^2(M) \to \Lambda_2(M)$, that is induced from the isomorphism of $T^*(M)$ with $T(M)$ that $g$ defines, and the Poincaré duality isomorphism $\#\colon \Lambda_2(M) \to \Lambda^2(M)$ that is defined by the assumption that $T(M)$ is orientable and a choice of unit-volume element $V$ has been made. With these assumptions, one can set $* = \# \cdot (g \wedge g)$.

It is illuminating to express these linear maps in matrix form. We restrict ourselves to the case in which $M$ is Minkowski space, i.e., $\mathbb{R}^4$ with the Minkowski scalar product $\eta$, which is true at least locally on any Lorentzian manifold. Instead of dealing with the (trivial) vector bundles $\Lambda_2(\mathbb{R}^4)$ and $\Lambda^2(\mathbb{R}^4)$ of bivector fields and 2-forms, respectively, we shall simply deal with their fibers, the vector spaces $A_2(\mathbb{R}^4) = \mathbb{R}^4 \wedge \mathbb{R}^4$ and $A^2(\mathbb{R}^4) = \mathbb{R}^{*4} \wedge \mathbb{R}^{*4}$ of bivectors and second-degree exterior forms, respectively.

Relative to the choice of framing for the two vector spaces $A_2(\mathbb{R}^4)$ and $A^2(\mathbb{R}^4)$ that we made in the previous section, the matrices for the three isomorphisms $\eta \wedge \eta$, $\#$, and $*$ take the form:

$$[\eta \wedge \eta]^{IJ} = \left[\begin{array}{c|c} -I & 0 \\ \hline 0 & I \end{array}\right], \quad [\#]_{IJ} = \left[\begin{array}{c|c} 0 & I \\ \hline I & 0 \end{array}\right], \quad [*]_J^I = \left[\begin{array}{c|c} 0 & I \\ \hline -I & 0 \end{array}\right]. \tag{3.1}$$

The last expression clearly shows the character of $*$ as defining a complex structure on the vector space $A^2(\mathbb{R}^4)$.

The basic plan for pre-metric electromagnetism [**4, 5**] is to replace the Hodge dual isomorphism with an isomorphism that is not defined in terms of any metric. For this, one need only replace the isomorphism $g \wedge g$ with a linear electromagnetic constitutive law $\kappa \colon \Lambda^2(M) \to \Lambda_2(M)$ for the medium.

The pre-metric form of Maxwell's equations is then:

$$dF = 0, \qquad \delta\mathfrak{h} = \mathbf{J}, \qquad \mathfrak{h} = \kappa(F), \tag{3.2}$$

in which we have defined $\delta = \#^{-1}d\#$. This illustrates the fact that the divergence operator is more intrinsically related to the definition of a unit-volume element than it is to the definition of a metric.

## 4 Linear electromagnetic constitutive laws [6-11]

Constitutive laws, whether mechanical or electromagnetic, represent a mathematical encapsulation of the empirical properties of a medium that relate to the coupling of inputs to responses. For instance, in the mechanical case, the input is a stress and the response is a strain. In the electromagnetic case, the inputs are the electric and magnetic field strength covector fields and the response is the electric displacement and magnetic flux density vector fields, which include contributions from the electric and magnetic dipoles that form in the medium. More generally, one associates the Minkowski field strength 2-form $F$ with the electromagnetic excitation bivector field $\mathfrak{h}$. In any case, one assumes that this association is invertible.

In its most general form, a (non-dispersive) electromagnetic constitutive law is supposed to take 2-forms on spacetime to bivector fields in a manner that defines a diffeomorphism $\kappa_x$: $\Lambda^2_x(M) \to (\Lambda_2)_x(M)$ for each $x \in M$. This brings us to our first specialization of the definition: We shall consider only the linear case; i.e., the case in which the fiber diffeomorphisms $\kappa_x$ are linear isomorphisms. Although the more nonlinear general case is incontestably important for the applications to nonlinear optics and possibly even quantum electrodynamics, nevertheless, there is enough work to do for the moment in simply establishing the physical basis for the mathematical methods in the linear case.

A second specialization is to non-conducting media. For such media, there is no coupling of an electric current with the electric field in the medium, as with Ohm's law, for instance, which says that for a linear conductive medium there is a current $\mathbf{J}_e = \sigma \mathbf{E}$ that flows in the medium in response to the applied electric field $\mathbf{E}$. This has the effect of changing the nature of the Maxwell equations by the introduction of a source term that is dependent upon the electric field.

A third specialization that we have made implicitly is to non-dispersive media. That is, we assumed that the constitutive map takes fibers of one bundle to fibers of the other one in an instantaneous and local way. One could also treat a linear electromagnetic constitutive law as a linear operator from 2-forms to bivector fields that might have a non-instantaneous or non-local character, such as an integral operator:

$$\mathfrak{h}(\mathbf{x}) = \int_M \kappa(\mathbf{x}, \mathbf{x}')[F(\mathbf{x}')]\, dV' \tag{4.1}$$

with a two-point (matrix-valued) tensor field $\kappa(\mathbf{x}, \mathbf{x}')$ as its kernel. However, since the physical nature of dispersion seems to suggest the possibility of taking the Fourier transform of the electromagnetic fields in question, and such matters are quite involved on a manifold that does not have an affine structure – or at least a homogeneous space structure – we shall address dispersive media in some later investigation. Similarly, we shall restrict ourselves to real matrices, since the imaginary contributions also seem to mostly relate to dispersion and the complex nature of the frequency-wave number domain.

Furthermore, since the nature of the results in this article is that they are most easily established by elementary matrix computations, we are restricting ourselves to the manifold $M = \mathbb{R}^4$, which amounts to considering a single fiber of each bundle $\Lambda^2(\mathbb{R}^4)$ and $\Lambda_2(\mathbb{R}^4)$. This has the effect of making the linear electromagnetic constitutive laws in question take the form of linear isomorphisms $\kappa: A^2(\mathbb{R}^4) \to A_2(\mathbb{R}^4)$. Locally, this only implies that the components of the matrices that we are dealing with are real constants, not functions on $M$, which amounts to assuming homogeneity of the constitutive law, but since we will not be differentiating the matrices, for the present purposes, the restriction is superfluous.

In summary, we have restricted ourselves to linear, non-conducting, non-dispersive electromagnetic media. It is not necessary, at this point, to specify whether the medium is vacuous or material – in the sense of the vanishing or non-vanishing of a mass density, – or isotropic or anisotropic.

For the previous choice of frame on $A_2(\mathbb{R}^4)$ and coframe on $A^2(\mathbb{R}^4)$ a linear electromagnetic constitutive law can be represented by an invertible $6 \times 6$ real matrix, which we express in block form:

$$\kappa^{IJ} = \left[ \begin{array}{c|c} -\varepsilon^{ij} & \gamma^{ij} \\ \hline -\hat{\gamma}^{ij} & (\mu^{-1})^{ij} \end{array} \right]. \tag{4.2}$$

Note that the negative signs appear as a result of the fact that we are coupling covariant tensors to contravariant ones in a manner that must be consistent with the Minkowski scalar product that we are replacing. In fact, without them, there would be no hope of deriving an almost-complex structure, which is not associated with the Hodge isomorphism for a positive-definite metric.

If the Minkowski field strength 2-form is represented in the form:

$$F = E_i \, b^i + B_i \, b^{i+3} = E_i \, \theta^0 \wedge \theta^i + \varepsilon_{ijk} \, B^i \, \theta^j \wedge \theta^k, \tag{4.3}$$

and the electromagnetic excitation bivector field is represented in the form:

$$\mathfrak{h} = D^i \, \mathbf{b}_i + H^i \, \mathbf{b}_{i+3} = D^i \, \mathbf{e}_0 \wedge \mathbf{e}_i + \varepsilon^{ijk} \, H_i \, \mathbf{e}_j \wedge \mathbf{e}_k, \tag{4.4}$$

then we can describe the linear map in its three-dimensional vectorial form as:

$$\mathbf{D} = -\varepsilon(E) + \gamma(B) \tag{4.5a}$$
$$\mathbf{H} = -\hat{\gamma}(E) + \mu^{-1}(B). \tag{4.5b}$$

Hence, we see that the 3×3 submatrix $\varepsilon$ represents the electric permittivity of the medium and the $\mu^{-1}$ submatrix is the inverse of its magnetic permeability.

This general case of a linear electromagnetic constitutive law as defined by (4.2) or (4.5a, b) is often referred to as defining a *bi-anisotropic* electromagnetic medium [**8-11**].

Interestingly, it seems to be an open question how one might realize this sort of medium with real-world electromagnetic materials (cf., [**11**]).

The matrices $\gamma$ and $\hat{\gamma}$ are sometimes called "magneto-electric" coupling terms. In the real case, they are most commonly associated with the Fresnel-Fizeau effect, in which the relative motion of a material medium couples to the *B* field and *E* field present in it to produce additional electric and magnetic polarization, respectively, much as the relative motion of electric charges produces magnetic fields in the frame of the observer. To first order in *v/c*, if the relative velocity of a spatially isotropic medium (defined below) is described by the spatial vector $\mathbf{v} = v^i \mathbf{e}_i$ then the magneto-electric matrices take the form:

$$\gamma = \hat{\gamma} = \left(\frac{\varepsilon\mu - 1}{\mu c}\right) \begin{bmatrix} 0 & -v^3 & v^2 \\ v^2 & 0 & -v^1 \\ -v^3 & v^1 & 0 \end{bmatrix}, \qquad (4.6)$$

in which the bracketed components can also be described as the matrix ad(**v**) of the adjoint map of $\mathbb{R}^3$, with the Lie algebra structure that it gets from the vector cross product; i.e., ad(**v**)(**w**) = **v** × **w**. In particular, the matrix $\gamma$ is anti-symmetric.

Some common special types of constitutive maps can be easily described in terms of the submatrices. We shall mention the ones for which the medium is not in a state of relative motion to the observer, so $\gamma = \hat{\gamma} = 0$.

Slightly less general than the case of a bi-anisotropic medium is the case of a medium in which the electric and magnetic properties are anisotropic, but there is no magneto-electric coupling between them. Such a medium has a linear constitutive law with a matrix of the form:

$$\kappa^{IJ} = \begin{bmatrix} -\varepsilon^{ij} & 0 \\ \hline 0 & (\mu^{-1})^{ij} \end{bmatrix}, \qquad (4.7)$$

in which the non-zero submatrices are also symmetric. Such a medium might take the form of a crystalline medium that has both dielectric and magnetic properties.

A special case of this would include the *anisotropic dielectric* media, for which:

$$\kappa^{IJ} = \begin{bmatrix} -\varepsilon^{ij} & 0 \\ \hline 0 & \frac{1}{\mu}\delta^{ij} \end{bmatrix}, \qquad (4.8)$$

such as some optical media, and *anisotropic magnetic* media, for which:

$$\kappa^{IJ} = \begin{bmatrix} -\varepsilon\delta^{ij} & 0 \\ \hline 0 & (\mu^{-1})^{ij} \end{bmatrix}. \qquad (4.9)$$

One then comes to the *spatially isotropic* media, for which:

$$\kappa^{IJ} = \begin{bmatrix} -\varepsilon \delta^{ij} & 0 \\ \hline 0 & \dfrac{1}{\mu}\delta^{ij} \end{bmatrix}, \tag{4.10}$$

which include the vacuum and most optical media.

It is also possible that $\varepsilon = 1/\mu = \lambda$, a case that Lindell [8] refers to as *general isotropy*, for which:

$$\kappa^{IJ} = \lambda \begin{bmatrix} -I & 0 \\ 0 & I \end{bmatrix}. \tag{4.11}$$

Of the anisotropic media, one can also consider *uniaxial* dielectrics or magnetic media, for which $\varepsilon$ or $\mu^{-1}$ have one distinct eigenvalue, respectively. Lindell also defines *bi-isotropic* media, for which:

$$\kappa^{IJ} = \begin{bmatrix} -\varepsilon \delta^{ij} & \gamma \delta^{ij} \\ \hline -\hat{\gamma} \delta^{ij} & \dfrac{1}{\mu}\delta^{ij} \end{bmatrix}, \tag{4.12}$$

in which $\varepsilon, \mu, \gamma, \hat{\gamma}$ are all constants.

As far as the symmetry of the general $\kappa$ is concerned, one can polarize it into a sum of three matrices:

$$^{(1)}\kappa^{IJ} = \tfrac{1}{2}(\kappa^{IJ} + \kappa^{JI}) - \frac{1}{6}\mathrm{Tr}(\tilde{\kappa})[\#]^{IJ}, \tag{4.13}$$

which is symmetric and traceless, and which Hehl and Obukhov [4] call the *principal part* of $\kappa$; an anti-symmetric matrix:

$$^{(2)}\kappa^{IJ} = \tfrac{1}{2}(\kappa^{IJ} - \kappa^{JI}), \tag{4.14}$$

which they call the *skewon* part of $\kappa$; and a trace:

$$^{(3)}\kappa^{IJ} = \frac{1}{6}\mathrm{Tr}(\tilde{\kappa})[\#]^{IJ}, \tag{4.15}$$

which they refer to as the *axion* part of $\kappa$. Note that the trace part $^{(3)}\kappa^{IJ}$ involves the scalar product $\mathbf{V}(F, G)$ that is defined by the unit-volume element $\mathbf{V}$ and not the Euclidian one that is defined by the choice of frame.

One can express these matrices in terms of the submatrices of $\kappa$ relative to our chosen frame as defined in (4.2) as:

$$^{(1)}\kappa^{IJ} = \begin{bmatrix} -\varepsilon_S & \gamma_S \\ \gamma_S^T & \mu_S^{-1} \end{bmatrix} - {}^{(3)}\kappa^{IJ}, \qquad (4.16)$$

in which $\varepsilon_S$ is the symmetric part of $\varepsilon$, $\mu_S^{-1}$ is the symmetric part of $\mu^{-1}$, and:

$$\gamma_S = \frac{1}{2}(\gamma + \hat{\gamma}^T), \qquad (4.17)$$

along with:

$$^{(2)}\kappa^{IJ} = \begin{bmatrix} -\varepsilon_A & \gamma_A \\ -\gamma_A^T & \mu_A^{-1} \end{bmatrix}, \qquad (4.18)$$

in which the subscript $A$ refers to the anti-symmetric parts of the $\varepsilon$ and $\mu^{-1}$ matrices, with:

$$\gamma_A = \frac{1}{2}(\gamma - \hat{\gamma}^T), \qquad (4.19)$$

and:

$$^{(3)}\kappa^{IJ} = \frac{1}{6}\left(\sum_{i=1,2,3} \gamma^{ii} - \hat{\gamma}^{ii}\right)[\#]^{IJ}. \qquad (4.20)$$

Hence, the axion part vanishes in media for which the magneto-electric submatrices have the same trace, such as with no magneto-electric coupling or media for which the magneto-electric submatrices are anti-symmetric. In a medium with symmetric electric and magnetic properties in a state of relative motion, the principal part of $\kappa$ is due to the $\varepsilon$ and $\mu^{-1}$ matrices, the skewon part is due to the Fresnel-Fizeau contribution, and the axion part vanishes.

## 5 Reduction to an almost-complex structure

Relative to the frames chosen above, we express the matrix of $\kappa$ in block-matrix form (4.2). Hence, the composition $\tilde{\kappa} = \# \cdot \kappa$ has a matrix of the form:

$$[\tilde{\kappa}]_J^I = \begin{bmatrix} -\hat{\gamma} & \mu^{-1} \\ -\varepsilon & \gamma \end{bmatrix}, \qquad (5.1)$$

in which we have suppressed the submatrix indices, for brevity.

In order for $\tilde{\kappa}$ to be proportional to a complex structure on the vector space $\Lambda^2(M)$, as would be consistent with the Hodge * isomorphism, if one sets * = $(1/\lambda)\tilde{\kappa}$ then it is necessary and sufficient that there be a real scalar $\lambda$ such that $\tilde{\kappa}^2 = -\lambda^2 I$.

Now, squaring $\tilde{\kappa}$ gives:

$$[\tilde{\kappa}^2]_J^I = \left[\begin{array}{c|c} \hat{\gamma}^2 - \mu^{-1}\varepsilon & \hat{\gamma}\mu^{-1} - \mu^{-1}\gamma \\ \hline -\varepsilon\hat{\gamma} + \gamma\varepsilon & \gamma^2 - \varepsilon\mu^{-1} \end{array}\right]. \tag{5.2}$$

Setting the submatrices equal to the corresponding submatrices of $-\lambda^2 I$ gives the following set of conditions on $\kappa$ in order for it define a complex structure on the vector space $A^2(\mathbb{R}^4)$:

$$\mu^{-1}\varepsilon = \lambda^2 I + \hat{\gamma}^2, \tag{5.3a}$$
$$\varepsilon\mu^{-1} = \lambda^2 I + \hat{\gamma}^2, \tag{5.3b}$$
$$\gamma\mu = \mu\hat{\gamma}, \tag{5.3c}$$
$$\gamma\varepsilon = \varepsilon\hat{\gamma}. \tag{5.3d}$$

In the event that $\gamma = \hat{\gamma}$, (5.3a, b) imply that $\varepsilon$ must commute with $\mu$, and (5.3c, d) say that $\gamma$ must commute with both the $\varepsilon$ and $\mu$ matrices. If all of the submatrices are symmetric then this would imply that they are also simultaneously diagonalizable in that case; i.e., they would all have the same principal frames.

The conditions that we just derived are not independent. For instance, one can obtain (5.3a) from (5.3b) and (5.3c). Furthermore, if one solves (5.3b) for $\varepsilon$ and (5.3c) for $\hat{\gamma}$ then one sees that (5.3d) follows automatically. Hence, the necessary and sufficient conditions for $\tilde{\kappa}$ to be proportional to a complex structure on the vector space $\Lambda^2(M)$ can be written:

$$\varepsilon = \lambda^2\mu + \mu\hat{\gamma}^2, \tag{5.4a}$$
$$\hat{\gamma} = \mu^{-1}\gamma\mu. \tag{5.4b}$$

This is a system of 18 equations in 36 unknowns, so since the dimension of the space of independent solutions – namely, 18 – is equal to the number of components to the matrices $\mu$ and $\gamma$, one can choose them arbitrarily and define the matrices $\varepsilon$ and $\hat{\gamma}$ by means of these equations.

We should examine the extent to which the aforementioned conditions on the submatrices change under a change of frame. Although the group of linear transformations of $A^2(\mathbb{R}^4)$ that preserve $\tilde{\kappa}$ – up to a scalar constant – is seen to be $\mathbb{R}^+ \times GL(3; \mathbb{C})$, in which the factor $\mathbb{R}^+$ ( = positive real numbers) accounts for the possible change in $\lambda$ to $\lambda'$, the transformations of this group will generally not preserve the individual submatrices. One sees that the conditions (5.4a, b) are only invariant under transformations of the three-dimensional real and imaginary subspaces [2] of $A^2(\mathbb{R}^4)$ by elements of $GL(3; \mathbb{R})$. Such a transformation $A \in GL(3; \mathbb{R})$ would then take $\varepsilon$ to $\varepsilon' = A^{-1}\varepsilon A$, etc., and the conditions (5.4a, b) for the transformed submatrices are altered only by primes. Of course, this is to be expected, since if we had a Lorentzian structure to begin with, although the proper orthochronous Lorentz group gets represented by an $SO(3; \mathbb{C})$ subgroup of $GL(3; \mathbb{C})$, which then preserves $\tilde{\kappa}$, one does not expect the nature of the submatrices to be Lorentz-invariant (e.g., Fresnel-Fizeau effect).

## 6 Almost-complex linear electromagnetic constitutive laws

Now, suppose that we are dealing with a linear electromagnetic constitutive law $\tilde{\kappa}$ on $A^2(\mathbb{R}^4)$ whose square is proportional to $-I$, which we refer to as an *almost-complex linear electromagnetic constitutive law*. Hence, its matrix (4.2) relative to our ongoing choice of basis satisfies the conditions (5.4a, b). Such a medium can be seen to be a special case of what Lindell [**8, 12**] calls a "self-dual" medium, for which $\hat{\kappa}$ is unchanged (up to a scalar factor) by a duality rotation of the form $\cos\theta\, I + \sin\theta\, *$, where $\theta$ is a choice of angle. However, this presupposes the existence of the Hodge duality isomorphism, which contradicts the spirit of pre-metric electromagnetism.

We now examine some special cases of submatrices $\varepsilon$, $\mu$, $\gamma$, $\hat{\gamma}$ that satisfy conditions (5.4a, b).

First, we consider media for which $\gamma = \hat{\gamma} = 0$. We see that the remaining condition (5.4a) says that one must have:

$$\varepsilon = \lambda^2 \mu. \tag{6.1}$$

Note that although this is the case for spatially isotropic media, such as the vacuum, it not a property of such a common medium as the anisotropic optical dielectric.

If both the $\varepsilon$ and $\mu$ are simultaneously diagonalizable – i.e., they have the same principal frame – then if we represent the two matrices as:

$$\varepsilon = \begin{bmatrix} \varepsilon_1 & 0 & 0 \\ 0 & \varepsilon_2 & 0 \\ 0 & 0 & \varepsilon_3 \end{bmatrix}, \quad \mu = \begin{bmatrix} \mu_1 & 0 & 0 \\ 0 & \mu_2 & 0 \\ 0 & 0 & \mu_3 \end{bmatrix} \tag{6.2}$$

we can express the condition (6.1) by saying that:

$$\lambda^2 = \frac{\varepsilon_1}{\mu_1} = \frac{\varepsilon_2}{\mu_2} = \frac{\varepsilon_3}{\mu_3}. \tag{6.3}$$

The principal indices of refraction become:

---

[2] By "real" and "imaginary" subspaces, we mean the ones that are spanned by $b_i$ and $*b^i$, respectively. We shall not elaborate here, but only refer the curious to other work of the author [**13**].

$$n_i = \sqrt{\varepsilon_i \mu_i} = \lambda \mu_i = \frac{\varepsilon_i}{\lambda}, \tag{6.4}$$

so we see that this sort of medium does not have to have the same speed of propagation in all directions, as one might guess.

From a phenomenological standpoint, the requirement that the matrices have the same principal frame seems somewhat restricting, since it would imply that the angular orientations of the electric dipoles that form in response to $E$ must be highly correlated with the angular orientations of the magnetic dipoles that form in response to $B$.

As for the cases in which the magneto-electric coupling is non-trivial, if we wish that $\gamma = \hat{\gamma}$ it is necessary that $\gamma$ commute with $\mu$. For instance, one could set $\hat{\gamma}^j_i = \gamma^j_i = \gamma \delta^i_j$, in which $\gamma$ is a scalar; this would give a special case of a bi-anisotropic medium. The $\varepsilon$ matrix would then take the form:

$$\varepsilon = (\lambda^2 + \gamma^2)\mu. \tag{6.5}$$

This also implies that $\varepsilon$, $\mu$, and $\gamma$ must have the same principal frame if they are symmetric.

As an example of an almost-complex medium for which $\gamma \neq \hat{\gamma}$, consider the general bi-isotropic medium. For such a medium, the conditions on the submatrices reduce to scalar equations that must be satisfied by the scalars $\varepsilon$, $\mu$, $\gamma$, $\hat{\gamma}$ that define them.

One way of seeing how an almost-complex structure $*$ on $\Lambda^2(\mathbb{R}^4)$ is equivalent to a conformal class of Lorentzian structures is to note that since we have already pointed out that a Lorentzian structure defines an almost-complex structure on by way of the Hodge $*$ operator, all one must do is to show how the isomorphism $*$ induces a light cone in each tangent space to $\mathbb{R}^4$; we shall use the tangent space at the origin as typical. The most elegant way to define lightlike lines in that tangent space in terms of 2-forms is to use the projective geometric notions that any 2-plane in $\mathbb{R}^4$ is associated with an equivalence class of decomposable 2-forms that differ by a non-zero scalar multiple (Plücker-Klein embedding) and that the exterior product of two such decomposable 2-forms vanishes iff the 2-planes that they represent intersect non-trivially. Now, a 2-form $F$ is decomposable iff $F \wedge F = 0$, which defines the five-dimensional Klein quadric in $\Lambda^2(\mathbb{R}^4)$. If one further restricts oneself to decomposable 2-forms for which $F \wedge *F = 0$ then one is considering 2-forms with the property that the 2-planes defined by $F$ and $*F$ intersect non-trivially. Since they cannot be identical, the intersection must be a line through the origin in $\mathbb{R}^4$.

Now look at the set of all lines through the origin in $\mathbb{R}^4$ that correspond to 2-forms $F$ that satisfy the pair of quadratic equations $F \wedge F = F \wedge *F = 0$, which one can think of as:

$$V(F, F) = \kappa(F, F) = 0, \tag{6.6}$$

in the event that $*$ is associated with a constitutive law; we shall call such 2-forms *isotropic*. In order to see that this set of lines defines an actual light cone, it helps to know that one can define a complex orthogonal structure on $\Lambda^2(\mathbb{R}^4)$ by way of:

$$(F, G)_{\mathbb{C}} = \kappa(F, G) + i\, V(F, G). \tag{6.7}$$

Under the $\mathbb{C}$-linear isomorphism of $\Lambda^2(\mathbb{R}^4)$ with $\mathbb{C}^3$, given the standard Euclidian structure – which is an isometry, moreover – the set of all isotropic 2-forms corresponds to the complex quadric in $\mathbb{C}^3$ that is defined by:

$$\sum_{i=1,2,3} (Z^i)^2 = 0. \tag{6.8}$$

This algebraic surface is two-complex-dimensional – hence, four-real-dimensional – but by the homogeneity of the equation (6.8), it projects to a one-complex-dimensional – hence, two-real-dimensional – algebraic curve in $\mathbb{C}P^2$. This curve can be seen to be diffeomorphic to a light cone, so the almost-complex structure is associated with a conformal class of Lorentzian metrics [3].

Whether or not the conditions imposed on $\kappa$ by the reduction to an almost-complex structure have any bearing on whether the medium will exhibit birefringence (double refraction) or not can only be established by Fresnel analysis, which is beyond the scope of the present study, but certainly needs to be addressed.

## 7 Discussion

As long one is mostly concerned with the idealized case of the vacuum, or some bi-anisotropic media, more generally, the conditions that we have derived are not particularly damaging to pre-metric electromagnetism. However, since they are not satisfied by the anisotropic dielectrics one must be careful about trying to reason by analogy when one starts with phenomena that are relevant to such media – e.g., optical phenomena in crystals – and attempts to derive consequences in purely mathematical terms.

It should be emphasized that it is not necessary for the electromagnetic constitutive map to be proportional to an almost-complex structure in order for one to formulate the field equations of pre-metric electromagnetism. This is because all that is required for the definition of the divergence of bivector fields is the volume element that defines Poincaré duality and all that is required in order for one to associate 2-forms with bivector fields is the constitutive map.

The point at which it becomes necessary for the constitutive map to define an almost-complex structure is when one wished to make contact with conventional electromagnetism again. Basically, the Lorentzian structure on the tangent bundle first appears in the form of a reduction of the bundle $GL(6;\mathbb{R})(\Lambda^2)$ of real 6-frames on $\Lambda^2(M)$ to the bundle of complex 3-frames $GL(3;\mathbb{C})(\Lambda^2)$, as defined by the almost-complex structure, to the bundle $SO(3;\mathbb{C})(\Lambda^2)$ of oriented complex orthogonal 3-frames when one defines a complex Euclidian structure on the $\Lambda^2(M)$ that is defined by the almost-complex structure (cf., [**13**]). Since the Lie group $SO(3; \mathbb{C})$ is isomorphic to the identity component of the Lorentz group – viz., the proper orthochronous Lorentz group $SO_0(3, 1)$ – one also has that the bundle $SO(3;\mathbb{C})(\Lambda^2)$ is isomorphic to the bundle $SO_0(3, 1)(M)$ of oriented, time-oriented Lorentzian frames in the tangent spaces to $M$.

---

[3] For more details on the way that complex projective geometry relates to 2-forms and conformal structure, cf., [**13**]. In Hehl and Obukhov [**4**], there is a detailed discussion of how one can construct a Lorentzian metric somewhat indirectly by a process that was defined by Schönberg and Urbantke in terms of a triplet of 2-forms; i.e., a frame for $\mathbb{C}^3$.

Physically, this means that connection between pre-metric and metric electromagnetism, as well as the presence of gravitation that one generally associates with the Lorentzian structure on spacetime, is crucially dependent upon the manner by which the electromagnetic medium supports the propagation of waves. Indeed, if one recalls that the usual conception of spacetime geometry is crucially dependent upon the way that space and time are connected by the universality of the speed of light, one sees that the electromagnetic structure of spacetime is more physically and mathematically fundamental than the gravitational structure.

Furthermore, the reduction to an almost-complex structure does not seem to be entirely necessary for the propagation of waves on the physical grounds that, after all, one can still do optics for the aforementioned anisotropic dielectric media with isotropic magnetic properties. Hence, the role of this reduction still seems somewhat unclear in the physical picture.

Perhaps, the dominance of the electromagnetic coupling constant over the gravitational one is another indication that gravitation is the shadow that electromagnetism casts upon spacetime geometry.